\documentclass[aps,pra,twocolumn,superscriptaddress,longbibliography]{revtex4-1}

\usepackage{color}
\usepackage{setspace}
\usepackage{graphicx}
\usepackage{mathtools}
\usepackage{verbatim}
\usepackage{amsmath}
\usepackage{caption}
\usepackage{appendix}
\usepackage{esint}
\usepackage{commath}
\usepackage{tensor}
\usepackage{stackengine}
\usepackage{xcolor,soul}
\usepackage{mathrsfs}
\usepackage{dcolumn}
\usepackage{bm}
\usepackage[english]{babel}
\usepackage{geometry}
\usepackage{lipsum}
\usepackage{amsfonts}
\usepackage{titlesec}
\usepackage{etoolbox}
\usepackage{float}
\usepackage[utf8]{inputenc}
\usepackage{booktabs}
\usepackage{tabularx}
\usepackage{colortbl}

\begin{document}

\title{Active Electric Dipole Energy Sources: Transduction via Electric Scalar and Vector Potentials}

\author{Michael E. Tobar}
\affiliation{Quantum Technologies and Dark Matter Labs, Department of Physics, University of Western Australia, 35 Stirling Highway, Crawley, WA 6009, Australia.}
\author{Raymond Y. Chiao}
\affiliation{School of Natural Sciences, University of California Merced, 5200 N. Lake Rd, Merced, CA 95343, USA}
\author{Maxim Goryachev}
\affiliation{Quantum Technologies and Dark Matter Labs, Department of Physics, University of Western Australia, 35 Stirling Highway, Crawley, WA 6009, Australia.}
\email{michael.tobar@uwa.edu.au}

\date{\today}

\begin{abstract}
The creation of electromagnetic energy may be realised by engineering a device with a method of transduction, which allows an external energy source, such as mechanical, chemical, nuclear etc., to be impressed into the electromagnetic system through a mechanism that enables the separation of opposite polarity charges. For example, a voltage generator, such as a triboelectric nanogenerator, enables the separation of charges through the transduction of mechanical energy, creating an active physical dipole in the static case, or an active Hertzian dipole in the time-dependent case. The net result is the creation of a static or time-dependent permanent polarisation respectively, without an applied electric field and with a non-zero vector curl. This system is the dual of a magnetic solenoid or permanent magnet excited by a circulating electrical current or fictitious bound current respectively, which supplies a magnetomotive force described by a magnetic vector potential and a magnetic geometric phase proportional to the enclosed magnetic flux. Thus, the active electric dipole voltage generator has been described macroscopically by a circulating fictitious magnetic current boundary source and exhibits an electric vector potential with an electric geometric phase proportional to the enclosed electric flux density. This macroscopic description of an active dipole is a semi-classical average description of some underlying microscopic physics, which exhibits emergent nonconservative behaviour not found in classical closed-system laws of electrodynamics. We show that the electromotive force produced by an active dipole in general has both electric scalar and vector potential components to account for the magnitude of the electromotive force it produces. Independent of the electromagnetic gauge, we show that Faraday's and Ampere's law may be derived from the time rate of change of the magnetic and dual electric geometric phases. Finally, we analyse an active cylindrical dipole in terms of scalar and vector potential and confirm that the electromotive force produced, and hence potential difference across the terminals is a combination of vector and scalar potential difference depending on the aspect ratio (AR) of the dipole. For long thin active dipoles (AR approaches 0) the electric field is suppressed inside, and the voltage is determined mainly by the electric vector potential. For large flat active dipoles (AR approaches infinity) the electric flux density is suppressed inside, and the voltage is mainly determined by the scalar potential.

\end{abstract}


\maketitle

\section{Introduction}

Classically  a permanent polarisation consists of equal and opposite charges, $\pm q_e^i$, displaced by a finite distance, $\vec{L}$, to create a macroscopic electric dipole moment (EDM), $\vec{d}=q_e^i\vec{L}$, where the vector direction is defined from $-q_e^i$ to $+q_e^i$ (with net charge $=0$). For an active system, the charges are displaced by an external impressed force per unit charge to seperate positive and negative charges in the induction process. This concept is the basis of generating electrical power from an external energy source, which supplies a non-conservative electromotive force \cite{Tobar2021,RHbook2012,Balanis2012,Volkakis2012}, allowing a voltage to exist across positively and negatively charged terminals. This means an external force in the opposite direction of the Coulomb force is required to keep the charges in static equilibrium, otherwise they will accelerate towards each other.  At large distances from the dipole, the electric field appears as an ideal dipole field determined by the EDM, $\vec{d}=q_e^i\vec{L}$. The ideal dipole exist only in the limit as $\vec{L}\rightarrow0$ and $q_e^i\rightarrow\infty$. In contrast, for distances close to the separated charges a dipole has more complex electromagnetic structure, and such non-ideal dipoles are commonly referred to as a ``physical" dipole.

\begin{figure}[t]
\includegraphics[width=1.0\columnwidth]{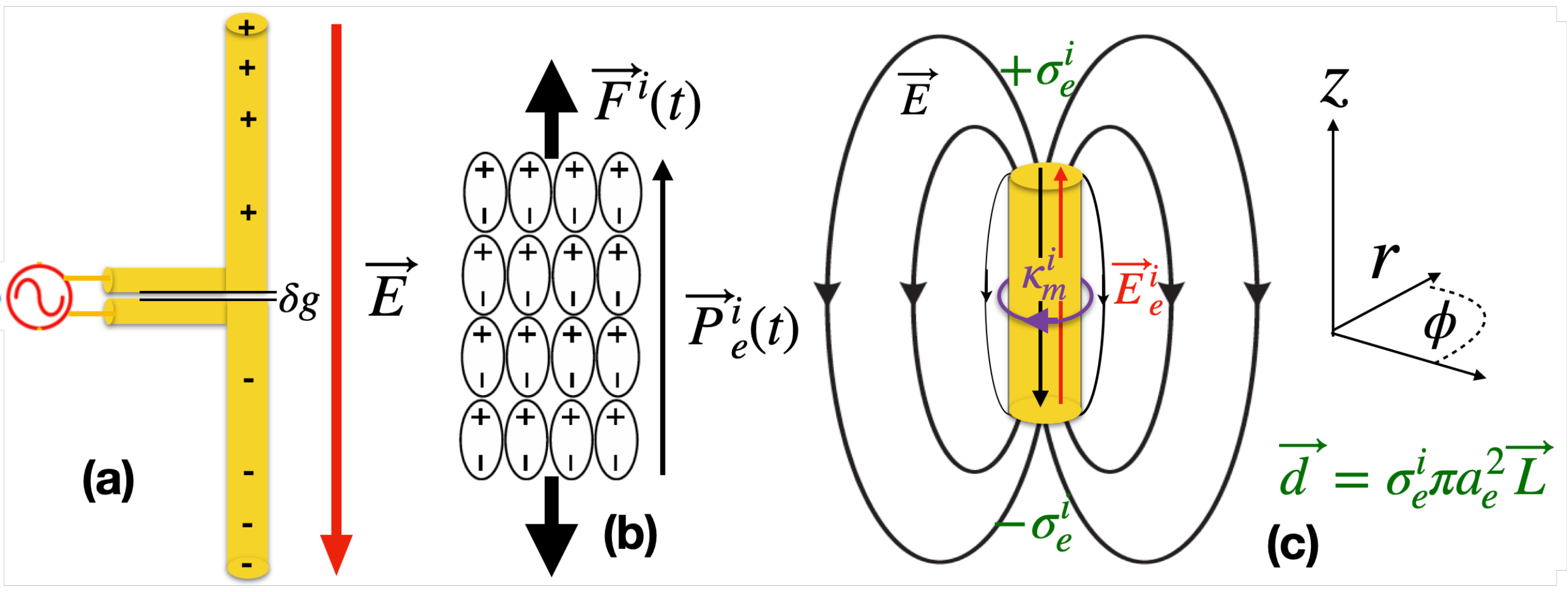}
\caption{(a) A free charge active Hertzian dipole antenna, (b) an active dipole bound charge nanogenerator. Both can be modelled by a voltage source with a capacitive output impedance \cite{Tobar2021}. (c) The equivalent macroscopic model of the active dipole with oppositely polarity surface charges, $\pm \sigma^i_{e}$, where $q_e^i=\sigma_e^i\pi a_e^2$, and  $a_e$, is the effective radius that the charge is spread over. The external force per unit charge, $\vec{E}_{e}^i=\frac{\vec{F}^i}{q_e^i}$, is finite and supplies the energy to seperate (and hence polarize) the impressed charges. The voltage output across the dipole can be modelled by an effective azimuthal magnetic surface current boundary source, which modifies Faraday's law, given by $\vec{J}_m^i=-\vec{\nabla}\times\vec{E}_{e}^i$. For a constant value of $\vec{E}_{e}^i$, the  effective magnetic current is on the radial surface so $\vec{\kappa}_m^i=\vec{J}_m^i\delta(r-a_e)$  (Weber convention for magnetic current). The separated free charges then generate a conservative electric field, $\vec{E}$, inside and outside the voltage source.}
\label{eleV2}
\end{figure}

The ideal oscillating time dependent active cylindrical dipole antenna is commonly known as a Hertzian dipole, and in the quasi static limit, $|\vec{r}|<\frac{\lambda}{6}$ ($\lambda$ is the wavelength of the radiation), the electrostatic near field dominates such that it is a maximum on the conducting cylindrical boundary as shown in Fig. \ref{eleV2} \cite{White1999}. Within the dipole, and active energy source drives the dipole through a gap spacing ($\delta g$), which is much smaller than the dipole dimensions, and thus the voltage and current oscillate out of phase as reactive power (no work is done) driven by an effective electric vector potential \cite{Tobar2021}. In contrast, external to the dipole, the electric field can be describe by either an electric scalar or vector potential, as the field is capable of doing work on a test charge, but also exist as a reactive near field (or fringing field) due to the unusual boundary condition between the outside and inside of an active electric dipole, where the tangential electrical field is maximum at the boundary \cite{Tobar2021,RHbook2012,Balanis2012,Volkakis2012,White1999}. Such active dipoles are usually configured with a balun, and can be used to generate or detect tangential electric fields, in particular they are used to characterise the near field of many systems, including antennas, materials and electrical fields in biological systems \cite{Smith1979,Wang2019,Baudry2007,Nobby2005,Jiang:2015wo,Yousefi:2015wu}.

In the case that the medium is an insulator a macroscopic active bound charge dipole is known as an electret \cite{Tobar2021} and exhibit a quasi-permanent polarisation (a metastable state), which can last for years) in the absence of an applied electric field. A common form of active electret is the nanogenerator \cite{WANG20179,WANG201774}, which are commonly used for energy harvesting and electricity generation \cite{WANG20179,WANG201774,Yang20122833,Ertuk2011,Jean-Mistral2012,mi11030267,Asanuma2013}. For example, many energy harvesting electret systems are based on triboelectric nanogenerators, where mechanical motion of the nanogenerator creats a time dependent polarisation, which is a displacement current. The standard Mawell-Faraday law cannot explain the emf produced as there is no significant variation of the net magnetic flux through the plane of the circuit \cite{Jenkins2020}. It was shown that electrons are being transported against the average electric field by a nonconservative force (or emf), effectively acting as a negative impedance through mechanical motion, and a microscopic quantum mechanical theory was developed to explain this effect  \cite{Jenkins2020}.  In this work we deal with the equivalent macroscopic theory that covers the general emf generator, but behind each generator or battery, there should be a similar microscopic theory based on emergent phenomena \cite{Jenkins2020,Alicki2020,Alicki2019,Hwang2012,LIU2020100066,Ilan:2020to}. In general any non-conservative generated emf can be explained by an impressed force per unit charge (the same unit of electric field), which creates a polarisation without an electric field, and in general has non-zero vector curl that can be inputted into Faradays law as forcing function with a corresponding fictitious magnetic current boundary source term as shown in Fig. \ref{eleV2}.

Furthermore, related to this, modern polarisation theory introduced in the 1990s \cite{Resta1992,Resta1994,King-Smith1993} has shown that the general definition of the polarisation was not solely calculable from bulk characteristics of the volume of bound charge, and that a change of polarisation only had physical meaning if it was quantified by using a geometric phase. This technique has been very successful in first-principles studies of spontaneous polarisation in ferroelectric materials (creation of a ferroelectret)\cite{RandV2007,Onoda2004,Asanuma2013}, it has also been shown that this emergent behaviour for a biaxial anisotropic photonic system may be explained using only classical electromagnetic concepts \cite{7820047}, and it was recently shown that a ferroelectric phase transition due to a soft phonon mode mode induced biaxial anisotropy in a perovskite material \cite{9523875}

In this work we use the fact that a permanent vector polarisation, generated without an electric field has both a non-zero curl and divergence. For the curl of the polarisation to be nonzero, an energy input is required to separate the bound charge; this describes a permanent electret or energy harvesting material \cite{Tobar2021,Sessler2016,Zi20152340,Wang2013,WANG20179,WANG2020104272,mi11030267,Gross62,Jean-Mistral2012,C6TA09590A} as well as the properties of ferroelectric domain walls \cite{Vasudevan:2017uf}. This description is also similar to an active dipole in antenna theory, a voltage source in circuit theory \cite{Tobar2021,RHbook2012,Balanis2012,Volkakis2012,White1999}, or an active dipole emitter in quantum theory \cite{Drezet2016,Drezet2017,Drezet17b}, where an external nonconservative force (sometimes referred as a fictitious or pseudo force) is described by an impressed electric field (some times referred as a fictitious or pseudo electric field) \cite{Pikulin16,Yu19,Ilan:2020to} with a nonzero curl (one could call this a polarisation). Furthermore, the electret, energy harvester, or ferroelectric domain may be classified as an active bound charge dipole. We may recognize this active dipole term generally as a nonconservative curl force term, which necessarily modifies Faraday's law, and is only present internally to the active antenna, voltage source, electret, or ferroelectric domain and not present globally outside the active device. As with all curl forces \cite{Berry_2013,Berry_2020,Berry2015,Guha:2020uh,SpinCurlF2013,Strange_2018}, this nonconservative term cannot be characterized by a scalar potential; on the other hand, it has been recently shown to be characterized via an electric vector potential \cite{Drezet2016,Drezet2017,TobarModAx19,TOBAR2020,Tobar2021,Tobar2022}, and we show that the permanent polarisation vector can be defined as a combination of a scalar and vector potential. Importantly, the electric vector potential gives a non-zero tangential surface term, which at the boundary can be viewed as an effective magnetic current \cite{Tobar2021}, an entity related to a geometric phase and a monopole instanton \cite{Song2021}. Furthermore, we find that the time rate of change of this electric geometric phase leads to the derivation of Ampere's Law (magnetomotive force), and the time rate of change of the well known magnetic Berry phase (or Aharonov-Bohm (AB), phase)\cite{AB1959} leads to the derivation of Faraday's law (electromotive force). This is consistent with prior work, which derives motive forces from the Aharonov-Bohm and Aharonov-Casher effects  \cite{Oh1994,ACEffect1984}.

\section{Quasi-Static Time Dependent Active Hertzian Dipole; Fields and Potentials}

For a dipole, some standard text book example assumes point charges, which are unphysical, a better approximation is to assume ideal surface charges, $\sigma_e^i$ \cite{Jackson99}, so $q_e^i=\sigma_e^i\pi a_e^2$ as shown in Fig.\ref{eleV2}c, so the electric force is spread over an area and solutions are non-divergent. Such permanent active electric dipoles occurs in bound charge (ideal electret) and free charge (battery, dipole antenna or electric generator) systems \cite{Tobar2021}. We thus may define the separation of free charge or bound charge by a polarisation vector as, 
\begin{align}
\vec{P}_e^i=\epsilon\vec{E}_e^i=\sigma_e^i\hat{z}, 
\label{Pbound}
\end{align}
where the polarisation vector is in the $\hat{z}$ direction, $\epsilon_r$ is the dielectric constant of any media involved, $\epsilon=\epsilon_0\epsilon_r$ and $\sigma_e^i$ represents impressed free or bound charge respectively. In these cases an effective magnetic current surface density exists, as shown in Fig.\ref{eleV2}c, at the radial boundary of the dipole and acts as a source term, which has been shown to be given by \cite{Tobar2021},
\begin{align}
\vec{\kappa}_m^i=-\frac{\sigma_e^i}{\epsilon}\hat{\phi},
\label{MagCurrent}
\end{align}
in the Weber convention and is in the azimuthal direction. Next we consider the general time dependent case.

Maxwell's equations for an ideal voltage generator with impressed bound or free charge ($\epsilon=\epsilon_0$) volume density, $\rho_e^i$, has been shown to be given by \cite{Tobar2021}  (Weber convention),
\begin{equation}
\vec{\nabla}\cdot\vec{E}=\frac{\rho_e^i}{\epsilon}~~\text{and}~~\vec{\nabla}\cdot\vec{E}_e^i=-\frac{\rho_e^i}{\epsilon},\label{Vs1}
\end{equation}
\begin{equation}
\vec{\nabla} \times \vec{B}-\epsilon\mu_0\frac{\partial \vec{E}}{\partial t}=\mu_0(\vec{J}_e^i+\vec{J}_f);~\vec{J}_e^i=\epsilon\frac{\partial\vec{E}_e^i}{\partial t},\label{Vs2}
\end{equation}
\begin{equation}
\vec{\nabla} \cdot \vec{B}=0,\label{Vs3}
\end{equation}
\begin{equation}
\vec{\nabla} \times \vec{E}+\frac{\partial \vec{B}}{\partial t}=0~~\text{and}~~\vec{\nabla} \times \vec{E}_e^i=-\vec{J}_m^i. \label{Vs4}
\end{equation} 
or in terms of the total electric field, $\vec{E}_T$ by
\begin{equation}
\vec{\nabla}\cdot\vec{E}_T=0,\label{Vss1}
\end{equation}
\begin{equation}
\vec{\nabla} \times \vec{B}-\epsilon\mu_0\frac{\partial \vec{E}_T}{\partial t}=\vec{J}_f,\label{Vss2}
\end{equation}
\begin{equation}
\vec{\nabla} \cdot \vec{B}=0,\label{Vss3}
\end{equation}
\begin{equation}
\vec{\nabla} \times \vec{E}_T+\frac{\partial \vec{B}}{\partial t}=-\vec{J}_m^i, \label{Vss4}
\end{equation}
with the following constitutive relations
\begin{equation}
 \vec{E}_T=\vec{E}_e^i+\vec{E}
 \label{freeE}
\end{equation}
Here, $\vec{J}_f$ in the lossless case has zero divergence, since $\rho_f=0$, and $\vec{J}_m^i$ also has zero divergence since $\rho_m^i=0$. $\vec{J}_{m}^i$ exists on the radial boundary of the dipole, and drives the impressed electric field, $\vec{E}^i_e$, by the left hand rule and also sets the boundary condition for the parallel components of the fields on the radial boundary. Here the $\frac{\partial \vec{B}}{\partial t}$ term in eqn.(\ref{Vss4}) can be identified as a magnetic displacement current and $\vec{J}_f$ can only exist if an external circuit is coupled to the ideal voltage generator\cite{Yang20122833,Ertuk2011,WANG201774,Jean-Mistral2012,mi11030267,Asanuma2013} or the generator is non-ideal with an effective internal resistance. 

The modified form of these equations means in general an electric vector potential, $\vec{C}$, can be introduced, along with the electric scalar potential, $V$, and the magnetic vector potential, $\vec{A}$. The possible existence of an electric vector potential and a magnetic scalar potential has been postulated to exist through the dual of Maxwell's equations being excited by magnetic monopoles and magnetic currents\cite{Cabibbo1962,Zwanziger1971,Singleton:1995dp,Singleton96,Keller2018,Rajantie2012,Mignaco2001} and is known as two-potential theory. Moreover, the electrical engineering community have also shown that the dual of Maxwell's equation may be excited by non-conservative electromagnetic systems or voltage generators \cite{Tobar2021,RHbook2012,Balanis2012}, without the need for monopoles to exist. Thus, from two-potential theory, and given there is no magnetic scalar field in the system we are describing, we may write the potential of the defined fields in eqns. (\ref{Vs1})-(\ref{freeE}) as,
\begin{equation}
\vec{B}=-\mu_0\frac{\partial\vec{C}}{\partial t}+\vec{\nabla}\times\vec{A}\label{Pot1}
\end{equation}
\begin{equation}
\vec{E}=-\vec{\nabla}V-\frac{\partial\vec{A}}{\partial t}\label{Pot2}
\end{equation}
\begin{equation}
\vec{E}_e^i=\frac{\vec{P}_e^i}{\epsilon}=\vec{\nabla}V-\frac{1}{\epsilon}\vec{\nabla}\times \vec{C}; \  \label{Pot3}
\end{equation}
\begin{equation}
\vec{E}_T=\frac{\vec{D}_T}{\epsilon}=-\frac{1}{\epsilon}\vec{\nabla} \times \vec{C}-\frac{\partial\vec{A}}{\partial t}\label{Pot4}
\end{equation}
Note, the field that experiences the ``pure'' vector potential is $\vec{E}_T=\frac{\vec{D}_T}{\epsilon}=\vec{P}_e^i/\epsilon+\vec{E}$, for both the free and bound system.

Inside the active dipole the polarisation field, $\vec{P}_e^i$, exists without any applied electric field, with both vector and scalar potential components, with the scalar component exactly equal and opposite to the scalar potential of the $\vec{E}$ field, consistent with eqn. (\ref{Vs1}). Meanwhile, $\vec{E}_e^i$ and  $\vec{E}_T$ have the same vector curl and thus the same component of electric vector potential, while satisfying the constitutive given by eqn. (\ref{freeE}). 

Outside the active dipole, $\vec{E}_e^i=0$, which means from eqn. (\ref{Pot3}), $\vec{\nabla}V_{out}=\frac{1}{\epsilon}\vec{\nabla}\times\vec{C}_{out}$ since the electric flux density and electric field intensity are equal outside the dipole ($\vec{E}_{out}=\vec{E}_{T_{out}}$ outside). This gives us two ways to describe the electric field or flux density outside the active dipole, i.e. with either an electric scalar or vector potential. In the quasi static limit the solution is dominated by the electrostatic near field of the dipole, which is reactive with the internal impressed current and voltage necessarily out of phase \cite{Tobar2021}. Thus, the electric flux density can be thought as a continuation of the same vector potential within the dipole, with the electric flux density given by the left hand rule, sourced from the magnetic current at the boundary, as shown in Fig.\ref{eleV2}. This dual description of the potential outside the active dipole is analogous to how a scalar magnetic potential is a useful quantity to describe the magnetic field outside a permanent magnet, highlighting that either a magnetic scalar or vector potential can be used. 

Now by substituting the fields given in eqns. (\ref{Pot1}) and (\ref{Pot4}) back into the electric and magnetic Gauss' law we obtain
\begin{equation}
\frac{\partial(\vec{\nabla}\cdot\vec{A})}{\partial t}=0;~~\frac{\partial(\vec{\nabla}\cdot\vec{C})}{\partial t}=0,
\label{TD}
\end{equation}
so the divergences of the vector potentials must be time independent. Then by substituting either (\ref{Pot2}) or (\ref{Pot3}) into Gauss' law, and using (\ref{TD}) we obtain,
\begin{equation}
\nabla^2V=-\frac{\rho_e^i}{\epsilon}
\end{equation} 
Substituting, (\ref{Pot1}) and (\ref{Pot4}) into Faraday's law we obtain,
\begin{equation}
\vec{\nabla}\times\vec{\nabla}\times\vec{C}+\mu_0\epsilon\frac{\partial^2\vec{C}}{\partial t^2}=\epsilon\vec{J}_m^i,
\label{MagCurPot}
\end{equation}
then by substituting, (\ref{Pot1}) and (\ref{Pot4}) into Ampere's law we obtain,
\begin{equation}
\vec{\nabla}\times\vec{\nabla}\times\vec{A}+\mu_0\epsilon\frac{\partial^2\vec{A}}{\partial t^2}=\mu_0\vec{J}_f.
\end{equation}
It is well known that there is more than one set of potentials that can generate the same fields, given that $\vec{\nabla}\times\vec{\nabla}\times\vec{C}=-\vec{\nabla}^2\vec{C}+\vec{\nabla}(\vec{\nabla}\cdot\vec{C})$ to simplify we chose the gauge where the divergence of the vector potentials are zero (Coulomb Gauge), so we obtain,
\begin{equation}
\Box ^2\vec{C}=-\epsilon\vec{J}_m^i,
\label{CPot}
\end{equation}
and
\begin{equation}
\Box ^2\vec{A}=-\mu_0\vec{J}_f,
\label{APot}
\end{equation}

Thus, we have successfully calculated the potentials in terms of the impressed sources, $\vec{J}_m^i$ and $\rho_e^i$ as well as any free current in the system, $\vec{J}_f$. For the lossless system with no load, $\nabla\cdot\vec{J}_f=0$. Note, that the impressed current, $\vec{J}_e^i=\epsilon\frac{\partial\vec{E}_e^i}{\partial t}$, in our presentation is not considered a source term, as it is described as  a non-dissipative polarisation current, which can either be from free or bound charge, impressed by the external force per unit charge, $\vec{E}_e^i$.

\section{Geometric Phase of an Active Electric Dipole}

The magnetic Aharonov-Bohm (AB) effect is a phenomenon where a charged particle's wave function is effected by the magnetic vector potential, $\vec{A}$, despite both the electric and magnetic field being zero \cite{AB1959}. Underlying this effect is the general concept of geometric or Berry phase  \cite{Berry1984} apparent in many areas of physics \cite{Wilczek1989} and not restricted to quantum mechanics, which includes optics \cite{Chiao1990,Lipson:90}, condensed matter physics \cite{Resta_2000,Xiao2007},  fluid mechanics \cite{Perrot:2019bb}, and so forth. Other related effects includes; 1) The Aharonov-Casher effect \cite{ACEffect1984,Cimmino1989,Elion1993,AC2006,Grosfeld2011}, which describes the effect of neutral particles with magnetic moments, effected by an isolated static positive or negative electric charge. The isolated electric monopole charge distribution creates an effective charge vector potential experienced by magnetic particles, and has been measured using magnetic flux vorticies \cite{Elion1993} or neutrons (with a dipole moment) \cite{Cimmino1989}. Like the AB effect the  charge vector potential associated with the Aharonov-Casher effect reveals a geometric phase in a charge-vortex interaction \cite{AC1997}; 2) The He-McKellar-Wilkens effect \cite{HeMcK93,Wilkens94}, dual to the Aharonov-Casher effect, which looks at the effect of neutral particles with EDMs induced by a magnetic monopole, and; 3) The dual Aharonov-Bohm (DAB) effect, which associates a Berry phase with a permanent polarisation (macroscopic collection of EDMs), such as that exhibited by an electret \cite{Dowling1999,SPAVIERI200313,Chen2013} or ferroelectet\cite{Onoda2004} due to an electric vector potential.

Since we have defined a macroscopic polarisation with respect to a 3D electric vector potential $\vec{C}$, we may equate this to a 3D Berry connection, with the Berry curvature the field given by eqn.(\ref{Pot4}), $\vec{D}_T=\epsilon\vec{E}_T=\vec{P}_e^i+\epsilon\vec{E}$. In fact, the electric dipole is dual to the magnetic dipole, which was used in the original AB thought experiment, so on this premises a dual electric effect should exist, and has been considered previously for an active dipole system \cite{Dowling1999,Chen2013,SPAVIERI200313}. In the strict sense of duality, the DAB experiment requires monopoles to measure the DAB effect. However, the DAB geometric phase should be equivalent to the known one discovered in the 1990s \cite{Vanderbilt2018,Xiao2007}, due to the spontaneous permanent polarisation of a ferroelectric \cite{Onoda2004}, or the permanent polarisation of an electret in general \cite{Resta1992,Resta1994,King-Smith1993,SPAVIERI200313,Onoda2004}, and a magnetic monopole was not necessary to prove the existence of this already widely accepted geometric phase.

First lets consider semi-classically the well known AB magnetic Berry phase of a long cylindrical electromagnetic solenoid (or permanent magnet), $\Delta \phi_{B_ {AB}}$, and with the use of eq. (\ref{Pot1}) we can show,
\begin{multline}
\phi_{B_ {AB}}=\frac{q}{\hbar} \oint_{\mathcal{P}} \vec{A} \cdot d\vec{l}=\frac{q}{\hbar}\int_{S}\nabla\times \vec{A}\cdot d\vec{S} \\
=\frac{q}{\hbar} \int_{S} \vec{B} \cdot d\vec{S}+\mu_0\frac{q}{\hbar}\int_{S}\frac{\partial \vec{C}}{\partial t}\cdot d\vec{S}.
\label{ABB}
\end{multline}
Here, the closed path, $\mathcal{P}$, of integration of the magnetic vector potential on the LHS of eqn. (\ref{ABB}) encloses the surface, $S$, in which the magnetic flux flows, with the first term on the RHS the static contribution to the AB geometric phase, while the second term adds the time dependent term. For the static case if we consider, $\mathcal{P}$ as the path at the mid point of the solenoid around the the electric current boundary, the minimum value of enclosed magnetic flux will be given by the flux quantum, $\Phi_0=h/(2e)$, so that $\int_{S} \vec{B} \cdot d\vec{S}=n \Phi_0$ for a superconducting system with $n$ Cooper pairs ($q=2e$). In contrast, for a normal conductor with free electrons ($q=e$), $\int_{S} \vec{B} \cdot d\vec{S}=2n \Phi_0$ (measured by Webb et. al. \cite{Webb1985}). Thus, in general the static AB phase in both the superconducting and normal conducting case is given by, $\phi_{B_ {AB}}=2n\pi$.

Now we consider in analogy the dual electric phase $\phi_{E_ {AB}}$, and with the use of eq.(\ref{Pot4}) we obtain,
\begin{multline}
\phi_{E_ {AB}}=\frac{1}{q} \oint_{\mathcal{P}} \vec{C}\cdot d\vec{l}=\frac{1}{q}\int_{S}\nabla\times C \cdot d\vec{S}\\
=\frac{1}{q} \int_{S} \vec{D}_T \cdot d\vec{S}+\frac{\epsilon}{q} \int_{S}\frac{\partial \vec{A}}{\partial t}\cdot d\vec{S}.
\label{ABE}
\end{multline}
Here, the closed path, $\mathcal{P}$, of integration of the electric vector potential on the LHS of eqn. (\ref{ABE}) encloses the surface, $S$, in which the electric flux flows. Thus, in analogy, the first term on the RHS gives the static dual geometric phase, while the second gives the general time dependent term. For the static case the geometric phase depends on the enclosed electric flux, $\Phi_E=\int_{S} \vec{D}_T \cdot d\vec{S}$, which for a path, $\mathcal{P}$, at the mid point of the magnetic current boundary, the minimum value should be equal to the quantum of electric charge, $q=e$, for a single electron system or, $q=2e$, for a paired electron system. These equations should be valid for both bound-charge and free-charge actively polarized systems.

Considering modern polarisation theory based on Berry phase, the definition of polarisation was developed through the microscopic crystal lattice surface and volume charge distributions. As discussed by Vanderbuilt \cite{Vanderbilt2018}, modern polarisation theory is based on the heuristic replacement of the position vector, $\vec{r}\rightarrow i\nabla_{\vec{k}}$, by the $\vec{k}$-derivative operator. Thus, Berry phase is considered in momentum space rather that position space, and the polarisation is quantised, so that $\vec{P}\rightarrow\vec{P}+\Delta \vec{P}_e^i$, corresponds to $\phi_{E_ {AB}}\rightarrow\phi_{E_ {AB}}+2\pi$ \cite{King-Smith1993,Vanderbilt2018}. In contrast, our approach  allows  us to relate the same quanta of polarisation to the electric Berry phase in position space. In a similar way, Onoda et.al. \cite{Onoda2004} have described the topological nature of polarisation and charge pumping \cite{Thouless83} in ferroelectrics using an analogy to magnetostatics, by introducing a vector field with a Berry phase as a linear response of the covalent part of polarisation, which has incorporated a generalization of the Born charge tensor. In principle this microscopic type of description should be equivalent to a semiclassical emergent macroscopic description of polarisation with a non-zero curl and an electric vector potential as introduced in this work.  A similar strategy has also been presented in \cite{Fang92,Song2021}, and suggests the magnetic current boundary source is an instanton, with a Berry phase, which carries non zero crystal momentum.

\section{Motive Force Equations from the Time Dependence of Geometric Phase}

Previously an equivalence between the Aharonov-Bohm effect of a solenoid and the Aharonov-Casher effect of a charged rod has been demonstrated, where the time-dependent Aharonov-Casher phase was shown to induce a motive force via the SU(2) spin gauge field \cite{Oh1994}. In a similar way to the time dependence Aharonov-Bohm effect derives Faraday's law, responsible for electromagnetic induction and the electromotive force (emf). Here we show that the time dependence of the dual electric phase derives Ampere's law, the equation responsible for magnetomotive force (mmf). 

First we consider the time rate of change of eqn.(\ref{ABB}) and combining it with (\ref{Pot1}) we obtain,
\begin{multline}
-\frac{1}{\epsilon} \oint_{\mathcal{P}} \nabla\times\vec{C}\cdot d\vec{l}- \oint_{\mathcal{P}}\vec{E}_T\cdot d\vec{l}=\\
\frac{\partial}{\partial t}\oint_{S} \vec{B} \cdot d\vec{S}+\mu_0\epsilon\oint_{S} \frac{\partial^2\vec{C} }{\partial t^2}\cdot d\vec{S},
\end{multline}
which becomes,

\begin{multline}
\mathcal{E}= \oint_{\mathcal{P}}\vec{E}_T\cdot d\vec{l}=-\frac{\partial}{\partial t}\oint_{S} \vec{B} \cdot d\vec{S}-\\ \frac{1}{\epsilon}\oint_{S}\big(\nabla\times\nabla\times\vec{C}+\mu_0\epsilon \frac{\partial^2\vec{C}}{\partial t^2}\big)\cdot d\vec{S}\\
=-\frac{\partial\Phi_B}{\partial t}-\oint_{S}\vec{J}_m^i\cdot d\vec{S}
\label{emf}
\end{multline}
Here, $\mathcal{E}$, is defined as the electromotive force (emf), then from eqn.(\ref{emf}) we obtain, 
\begin{align}
\mathcal{E}_T=-I_{m_{enc}}=-\oint_{S}\vec{J}_m^i\cdot d\vec{S}=\mathcal{E}+\frac{\partial\Phi_B}{\partial t},
\label{Faraday}
\end{align}
which is Faraday's law \cite{Tobar2021}. Here, $I_{m_{enc}}$ is the enclosed effective current boundary source, and $\mathcal{E}_T$,the voltage across a dipole or total $emf$. 

Next we consider the time rate of change of eqn.(\ref{ABE}) and combining it with (\ref{Pot4}) we obtain, 
\begin{multline}
-\frac{1}{\mu_0} \oint_{\mathcal{P}} \nabla\times\vec{A}\cdot d\vec{l}+\frac{1}{\mu_0} \oint_{\mathcal{P}}\vec{B}\cdot d\vec{l}=\\
\epsilon\frac{\partial}{\partial t}\oint_{S} \vec{E}_T \cdot d\vec{S}+\epsilon\oint_{S} \frac{\partial^2\vec{A}}{\partial t^2} \cdot d\vec{S},
\end{multline}
which becomes,
\begin{multline}
\mathcal{F}=\frac{1}{\mu_0} \oint_{\mathcal{P}}\vec{B}\cdot d\vec{l}=\epsilon\frac{\partial}{\partial t}\oint_{S} \vec{E}_T \cdot d\vec{S}+\\ \frac{1}{\mu_0}\oint_{S}\big(\nabla\times\nabla\times\vec{A}+\mu_0\epsilon \frac{\partial^2\vec{A}}{\partial t^2}\big)\cdot d\vec{S}\\
=\epsilon\frac{\partial}{\partial t}\oint_{S} \vec{E}_T \cdot d\vec{S}+\oint_{S}\vec{J}_f\cdot d\vec{S},
\label{mmf}
\end{multline}
which is the integral form of Ampere's law  \cite{Tobar2021}. Here, $\mathcal{F}$, is defined as the magnetomotive force (mmf), then by rearranging eqn.(\ref{mmf}) we obtain,
\begin{align}
\mathcal{F}_T=I_{f_{enc}}=\oint_{S}\vec{J}_f\cdot d\vec{S}=\mathcal{F}-\frac{\partial\Phi_E}{\partial t},
\label{mmf2}
\end{align}
Here, $\mathcal{F}_T=I_{f_{enc}}=N\times I$, for an electric coil (some times referred as an elctromagnet) is the enclosed electrical current boundary source of a magnetic dipole or inductor coil with N turns. This could also be delivered by a permanent magnet, which has a fictitious bound magnetic current, $\vec{J}_b$, due to the permanent magnetisation $\vec{M}$, where $\vec{J}_b=\nabla\times\vec{M}$, so $\mathcal{F}_T=I_{b_{enc}}=\oint_{S}\vec{J}_b\cdot d\vec{S}$.

\begin{figure}[t]
\includegraphics[width=0.9\columnwidth]{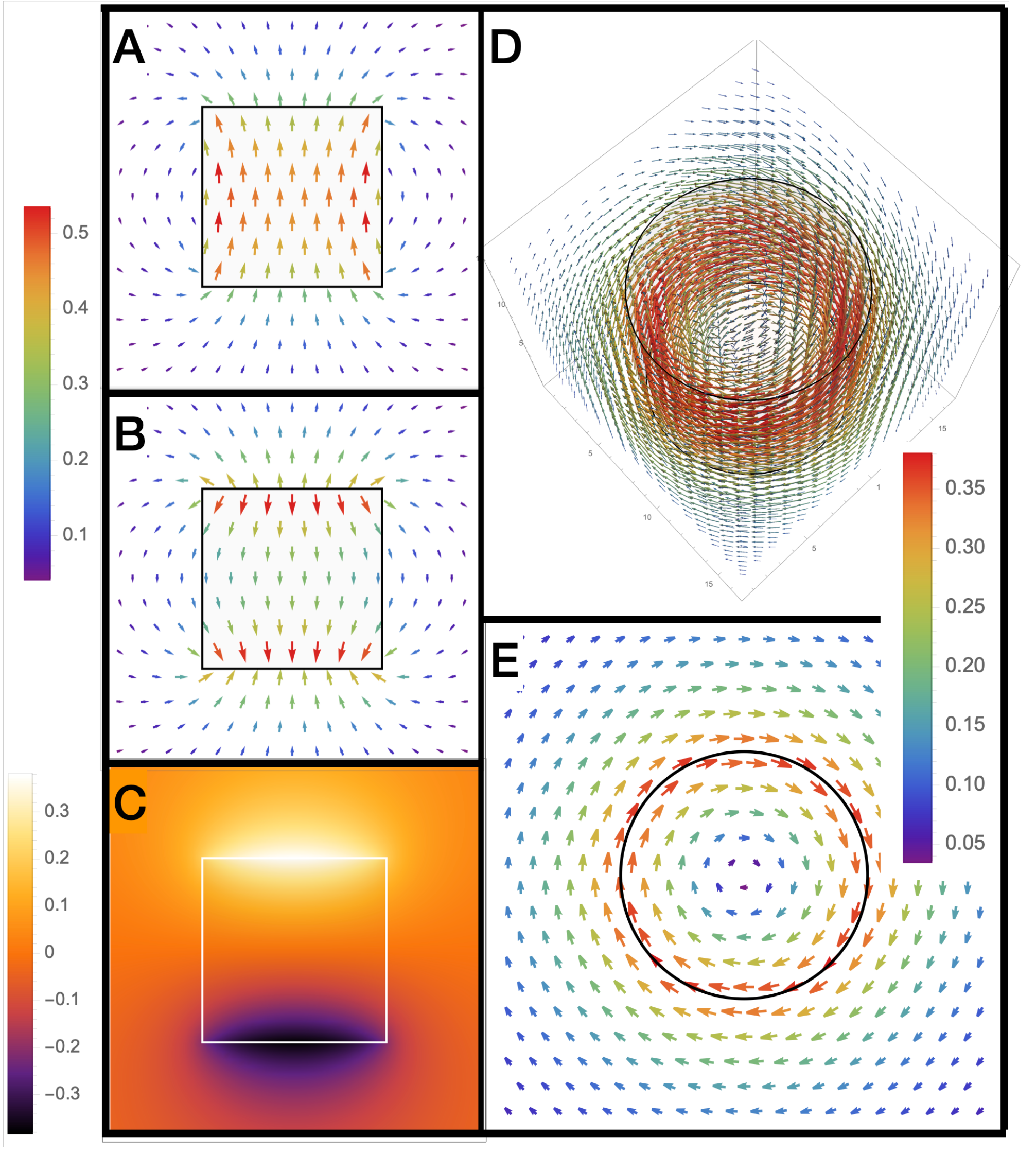}
\caption{Field and potential plots for a cylindrical dipole with AR=1. \textbf{A)} 2D vector plot of the normalized electric flux density $\frac{\vec{D}_T}{\sigma_e^i}$ at $y=0$, in the $(r-z)$ plane, calculated from eqns. (\ref{Dint}) and (\ref{Dint2}).  \textbf{B)} 2D vector plot of the normalized electric field $\frac{\epsilon\vec{E}}{\sigma_e^i}$ at $y=0$, in the $(r-z)$ plane, calculated from eqn. (\ref{Eint}). \textbf{C)} 2D colour density plot of the normalized electric scalar potential $\frac{\epsilon V}{\sigma_e^i}$ at $y=0$, in the $(r-z)$ plane, calculated from eqn. (\ref{Nscalarpot}). \textbf{D)} 3D vector plot of the normalized electric vector potential, $\frac{\vec{C}}{\sigma_e^i}$. \textbf{E)} 2D vector plot of the normalized electric vector potential, at $z=0$, in the $(r-\phi)$ plane, one can see the electric vector potential is maximum at the radial boundary where the magnetic current exists.}
\label{AR1}
\end{figure}
\begin{figure}[t]
\includegraphics[width=0.9\columnwidth]{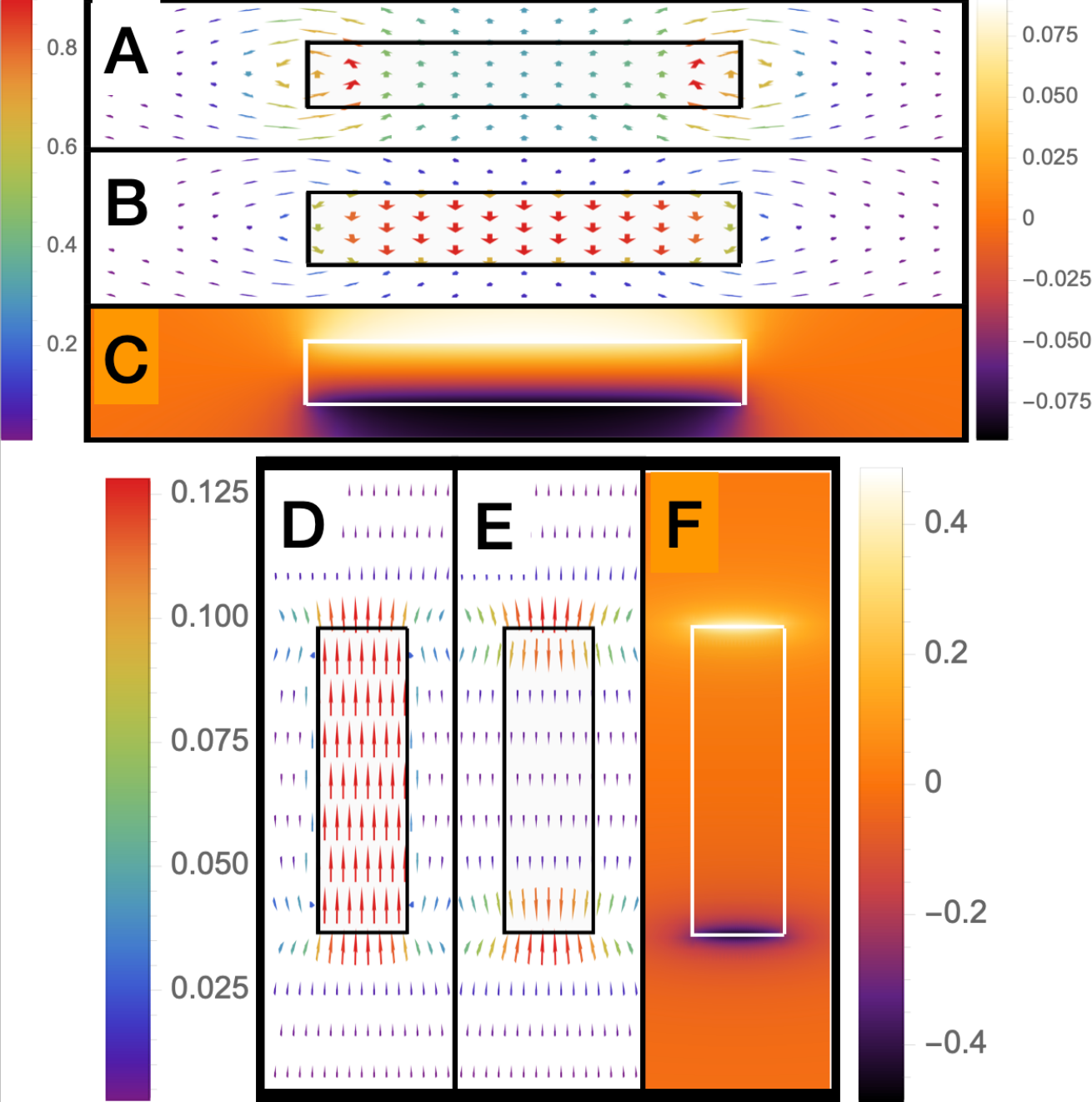}
\caption{Not to scale field and potential plots for a cylindrical dipole. \textbf{Above} AR=10: \textbf{A)} 2D vector plot of the normalized electric flux density $\frac{\vec{D}_T}{\sigma_e^i}$ at $y=0$, in the $(r-z)$ plane, calculated from eqns. (\ref{Dint}) and (\ref{Dint2}).  \textbf{B)} 2D vector plot of the normalized electric field $\frac{\epsilon\vec{E}}{\sigma_e^i}$ at $y=0$, in the $(r-z)$ plane, calculated from eqn. (\ref{Eint}). \textbf{C)} 2D colour density plot of the normalized electric scalar potential $\frac{\epsilon V}{\sigma_e^i}$ at $y=0$, in the $(r-z)$ plane, calculated from eqn. (\ref{Nscalarpot}). \textbf{Below} similar plots to above but with AR=0.1: \textbf{D)} $\frac{\vec{D}_T}{\sigma_e^i}$ at $y=0$, in the $(r-z)$ plane.  \textbf{E)} $\frac{\epsilon\vec{E}}{\sigma_e^i}$ at $y=0$, in the $(r-z)$ plane. \textbf{F)} $\frac{\epsilon V}{\sigma_e^i}$ at $y=0$, in the $(r-z)$ plane.}
\label{AR10}
\end{figure}
\begin{figure}[t]
\includegraphics[width=0.9\columnwidth]{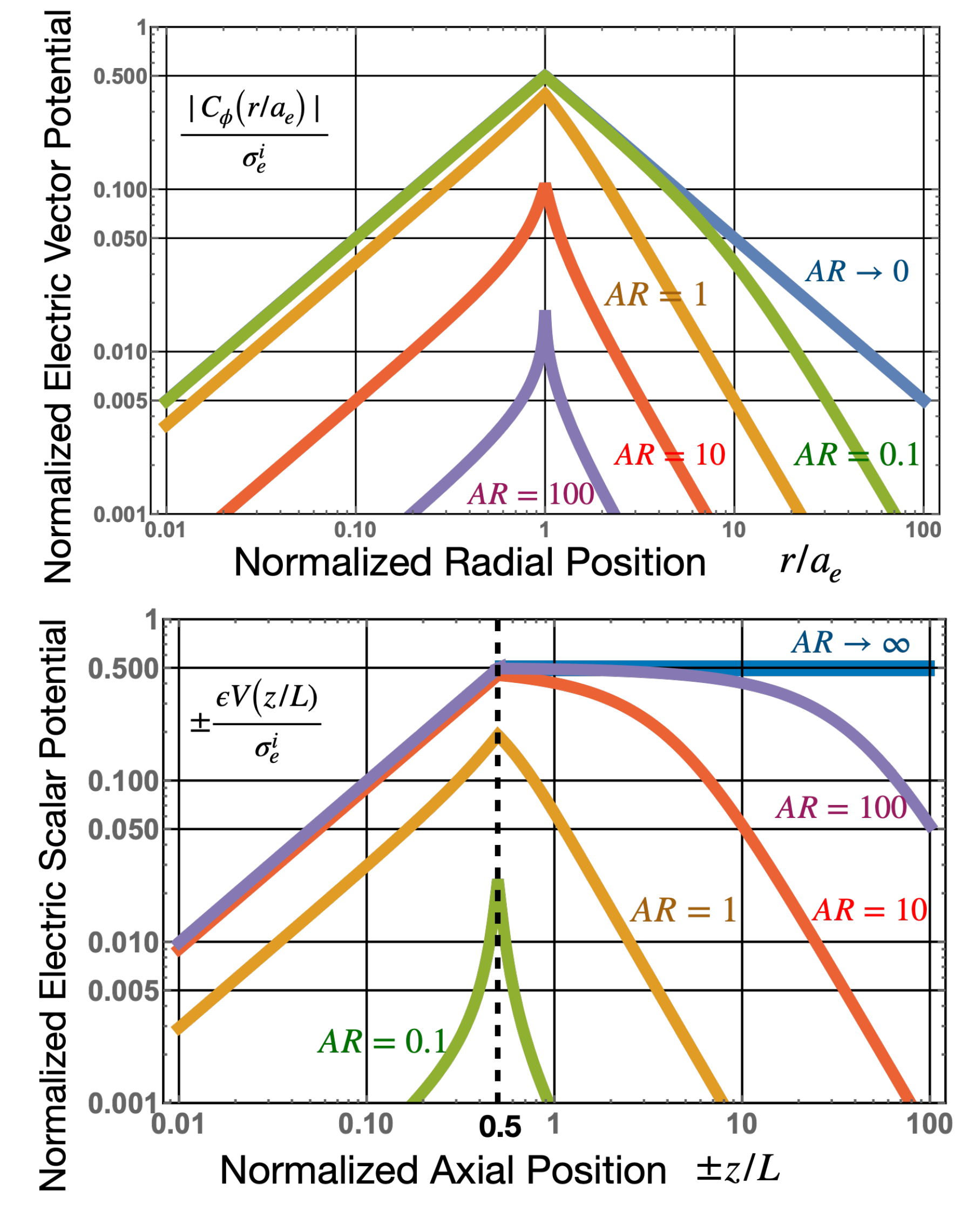}
\caption{Above: Normalized electric vector potential versus normalized radial distance, at $z=0$, from the centre of the electric dipole for various aspect ratios, compared to the infinitely long dipole ($AR\rightarrow0$). Below: Normalized electric scalar potential, versus normalized axial position, at $r=0$, from the centre of the electric dipole for various aspect ratios, compared to the infinely wide dipole $AR\rightarrow\infty$. Here, the length of the dipole is, $L$, where $AR=\frac{2a_e}{L}$, so the end face of the dipole are at $z/L=\pm\frac{1}{2}$.}
\label{VecP2}
\end{figure}
\begin{figure}[t]
\includegraphics[width=0.9\columnwidth]{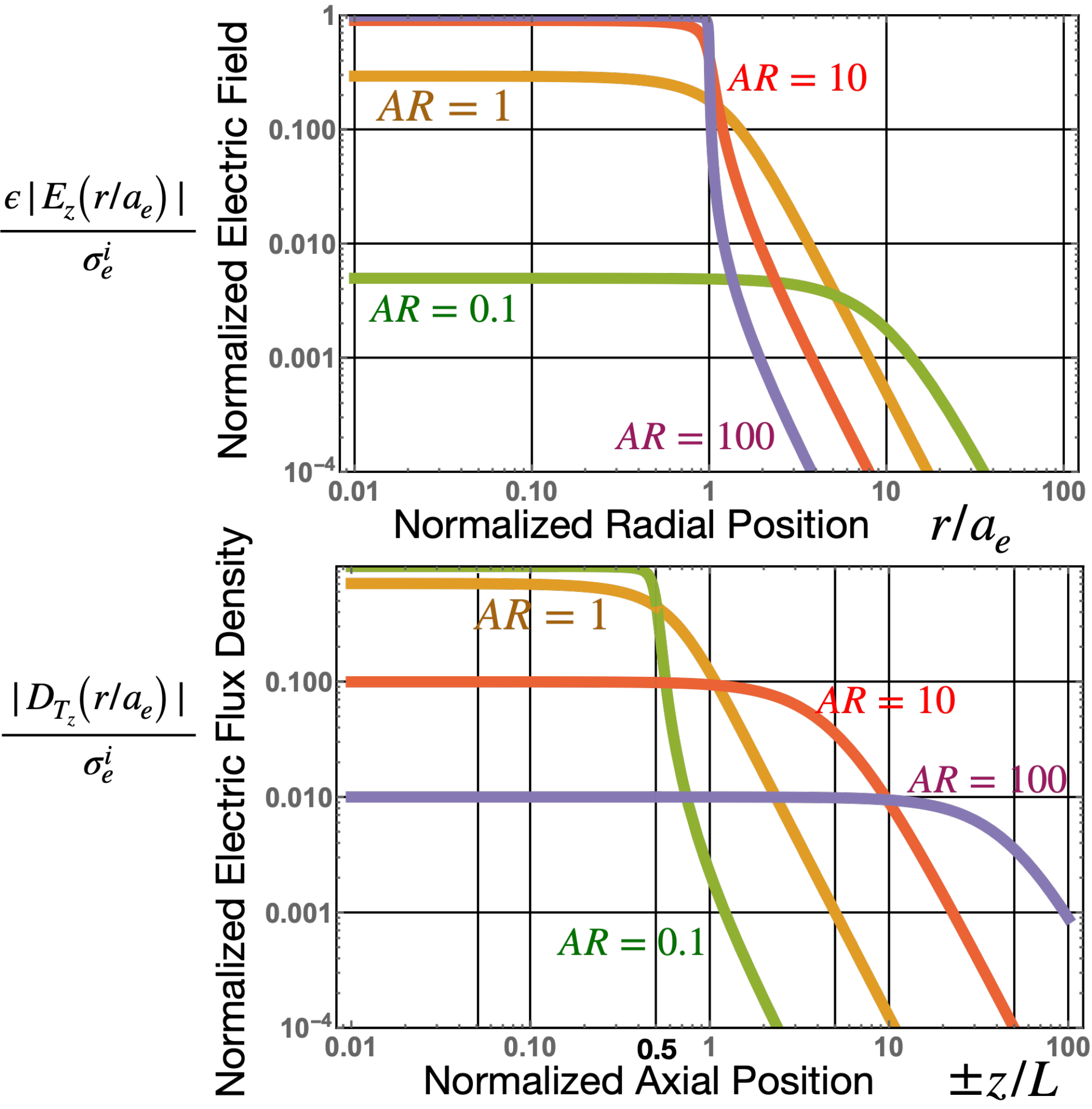}
\caption{Above: Normalized $z$ component of the electric field, $E_z$, versus normalized radial distance, at $z=0$, from the centre of the electric dipole for various aspect ratios. Note for the infinite dipole ($AR\rightarrow0$) the electric field is zero for all $r$. Below: Normalized $z$ component of the electric flux density, $D_{T_{z}}$, versus normalized axial distance, at $r=0$, from the midpoint of the electric dipole for various aspect ratios. Note, for the infinitely wide dipole ($AR\rightarrow\infty$) $D_{T_{z}}$ is zero for all $z$. Note the tangential $E_z$ field across the radial boundary of the dipole, at $\frac{r}{a_e}=1$, is continuous, while the normal $D_{T_{z}}$ field is continuous across the axial boundary at, $\frac{z}{L}=\pm\frac{1}{2}$.}
\label{Efield}
\end{figure}

\section{Electronic Properties of an Active Cylindrical Dipole}

In this section we analyze the electronic properties of a static (or quai-static) cylindrical active electronic dipole of varying aspect ratios ($AR=\frac{2a_e}{L}$), in terms of the fields and potentials as described in Section III. Here $L$ is the axial length, and $a_e$ the radius of the cylinder as shown in Fig.\ref{eleV2}. The aspect ratio was varied and the resulting electric scalar, $V$, and vector, $\vec{C}$, potentials, as well as the electric field, $\vec{E}$, and electric flux density, $\vec{D}$ were calculated, ranging from a flat pancake-like structure ($AR\rightarrow\infty$) to a long needle-like structure ($AR\rightarrow0$), with vector and density plots for some of these aspect ratios shown in Fig.\ref{AR1} and Fig.\ref{AR10}, while the values plotted against radial and axial positions are plotted in Fig.\ref{VecP2} and Fig.\ref{Efield}.

Assuming a constant impressed polarisation of $\vec{P}_e^i=\sigma_e^i\hat{z}$ within the boundaries of the active cylindrical dipole, a resulting constant impressed surface charge density will exist at each axial end face of, $\pm\sigma_e^i$, Correspondingly an impressed surface magnetic current density at the radial boundary ($r=a_e$) of value, $\epsilon\vec{\kappa}_m^i=-\delta(r-a_e)\sigma_e^i\hat{\phi}$ \cite{Tobar2021} will be present. The potentials and field can be calculated from the surface charge density and the surface magnetic current density using the following equations: \\
\\1) The electric scalar potential,
\begin{equation}
V(\vec{ r })=\frac{1}{4 \pi \epsilon}  \iint_{S}^ { \prime }  \frac{\sigma_e^i\left(\vec{r}^{\prime}\right) d A}{\left| \vec{ r }-\vec{ r } ^ {\prime} \right|},
\label{scalarpot}
\end{equation}
so the normalized value in cylindrical coordinates is given by,
\begin{multline}
\frac{\epsilon V(\vec{ r })}{\sigma_e^i}=\\
\frac{1}{4 \pi}  \int_{0}^ {a_e} \int _ {0}^ {2\pi} \frac{\delta(z^\prime-\frac{L}{2})-\delta(z^\prime+\frac{L}{2})}{\left| \vec{ r }-\vec{ r } ^ {\prime} \right|}r^\prime\mathrm{d}\phi^\prime\mathrm{d}r^\prime.
\label{Nscalarpot}
\end{multline}
2) The electric vector potential,
\begin{equation}
\vec{C}(\vec{ r })=\frac {\epsilon} {4\pi} \int _ {S}^ {\prime}  \frac{ \vec{\kappa}_{m}^i \left( \vec{r} ^ { \prime } \right) } { \left| \vec{ r }-\vec{ r } ^ {\prime} \right| } \mathrm{d}^{2} r^{\prime},
\end{equation}
so the normalized value in cylindrical coordinates is given by,
\begin{equation}
\frac{\vec{C}(\vec{ r })}{\sigma_e^i}=-\frac{a_e\hat{\phi}}{4\pi}\int_{-\frac{L}{2}}^ {\frac{L}{2}} \int _ {0}^ {2\pi}  \frac{\delta(r^\prime-a_e)} { \left| \vec{ r }-\vec{ r } ^\prime \right| } \mathrm{d}\phi^\prime\mathrm{d}z^\prime.
\end{equation}
3) The electric field vector ($\vec{E}=-\vec{\nabla}V$) ,
\begin{equation}
\vec{E}(\vec{ r })=\frac{1}{4 \pi\epsilon} \iint_{S}^ {\prime }\frac{\sigma_e^i\left(\vec{r}^{\prime}\right) d A}{\left(\vec{r}^{\prime}-\vec{r}\right)^{2}} \hat{\vec{r}}^{\prime} 
\end{equation}
so the normalized value in cylindrical coordinates is given by,
\begin{multline}
\frac{\epsilon\vec{E}(\vec{ r })}{\sigma_e^i}=\\
\frac{1}{4 \pi} \int_{0}^ {a_e} \int _ {0}^ {2\pi} \frac{\delta(z^\prime-\frac{L}{2})-\delta(z^\prime+\frac{L}{2})}{\left(\vec{r}^{\prime}-\vec{r}\right)^{2}} \hat{\vec{r}}^{\prime} r^\prime\mathrm{d}\phi^\prime\mathrm{d}r^\prime.
\label{Eint}
\end{multline}
4) The electric flux density ($\vec{D}=-\vec{\nabla}\times C$ ),
\begin{equation}
\vec{D}=-\frac{\epsilon}{4\pi}\int\frac{\vec{\kappa}_m^i\times (\vec{ r }-\vec{ r }^{\prime})}{ \left| \vec{ r }-\vec{r}^{\prime} \right|^3} dr^{\prime 2}
\label{Dint3}
 \end{equation}
 so the normalized value in cylindrical coordinates is given by,
\begin{multline}
\frac{\vec{D}(\vec{ r })}{\sigma_e^i}= \\
\frac{1}{4\pi}\int_{-\frac{L}{2}}^ {\frac{L}{2}} \int _ {0}^ {2\pi}  \frac{\delta(r^\prime-a_e)\hat{\phi}^{\prime}\times (\vec{ r }-\vec{ r }^{\prime})} { \left| \vec{ r }-\vec{ r } ^\prime \right|^3} \mathrm{d}\phi^\prime\mathrm{d}z^\prime.
\end{multline}
To verify this calculation we also used the relation, $\vec{D}_T(\vec{ r })=\epsilon\vec{E}_T(\vec{ r })+\vec{P}_e^i$, which leads to the following normalized values,
\begin{equation}
\frac{\vec{D}(\vec{ r })}{\sigma_e^i}=\frac{\epsilon\vec{E}(\vec{ r })}{\sigma_e^i}+\hat{z} \ \text{inside the dipole}
\label{Dint}
\end{equation}
\begin{equation}
\frac{\vec{D}(\vec{ r })}{\sigma_e^i}=\frac{\epsilon\vec{E}(\vec{ r })}{\sigma_e^i}\ \text{outside the dipole}
\label{Dint2}
\end{equation}
Both equations (\ref{Dint}) and (\ref{Dint2}), give the same result as (\ref{Dint3}) verifying our calculations.

\begin{figure}[t]
\includegraphics[width=0.9\columnwidth]{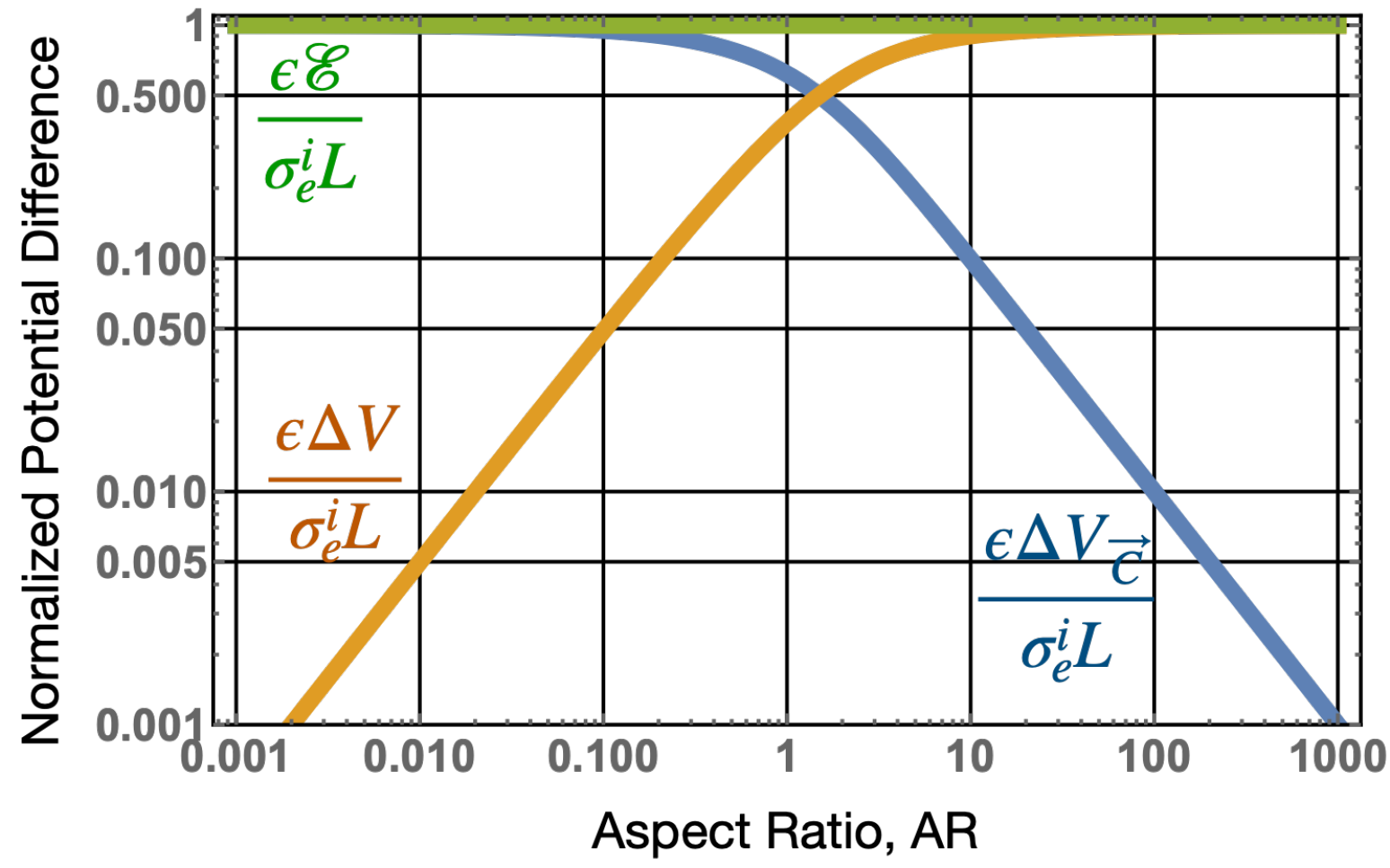}
\caption{Normalized potential difference across the terminals an active cylindrical dipole generator versus aspect ratio. The normalized $emf$ or voltage output, $\frac{\epsilon\mathcal{E}}{\sigma_e^iL}$, is shown in green and equal to unity independent of aspect ratio when setting the radius of the dipole to unity. The output voltage in general has both a scalar potential and vector potential as given by Eqns. (\ref{potdiff}) and (\ref{potdiff2}) and are plotted in orange and blue respectively. Results show that the vector potential dominates for small aspect ratios while the scalar potential dominates for large aspect ratios.}
\label{voltage}
\end{figure}

Some interesting points come out of these simulations, the potential difference ($\Delta V^i_e$ across the active dipole has both a scalar and vector potential component, and is equal to the electromotive force. Because $\vec{E}_e^i$ only exists within the active dipole source ($\vec{E}_e^i=0$ outside), the closed integral for the $emf$ can be replaced by a definite integral to give,
\begin{equation}
\Delta V^i_e=\mathcal{E}=\int_{-\frac{L}{2}}^{\frac{L}{2}}\vec{E}_e^i\cdot d\vec{l}=\Delta V_{\vec{C}}+\Delta V
\label{potdiff}
\end{equation}
where
\begin{equation}
\Delta V_{\vec{C}}=\int_{-\frac{L}{2}}^{\frac{L}{2}}\vec{E}_T\cdot d\vec{l}~~\text{and}~~\Delta V=-\int_{-\frac{L}{2}}^{\frac{L}{2}}\vec{E}\cdot d\vec{l}
\label{potdiff2}
\end{equation}
As shown in Fig.\ref{voltage}, as $AR\rightarrow0$, the $\pm\sigma_e^i$ charges will be separated by large distances when compared to the radius of the charge. In this case both $\vec{E}\rightarrow0$ and $\Delta V\rightarrow0$ (also see Fig.\ref{VecP2} and Fig.\ref{Efield}) so $\Delta V_{\vec{C}}$, is the main component of the voltage output. The opposite occurs for large aspect ratios for pancake-like structures. In this case the total electric field, $\vec{E}_T\rightarrow0$ or electric flux density $\vec{D}_T\rightarrow0$. For this case because $\vec{E}_e^i\approx-\vec{E}$ inside the dipole, and the potential difference between the axial end faces due to the scalar potential is equivalent to the $emf$ generated across the dipole, and $\Delta V_{\vec{C}}\rightarrow0$. This finding is consistent with \cite{Tobar2021}, which determined that the magnetic current boundary source best describes the output voltage of an AC or DC generator, rather than the electric field. Many authors assume $\vec{E}_T=0$, so under this assumption the $emf$ is only generated by a scalar potential, and is this were true, the near field of the active dipole in the quasi-static regime should be zero (screened). However, this is known to be generally not true, and it is well known that for long thin dipoles, such as an active dipole antenna, the near field is dominated by an electric field.

\section{Discussion}

A macroscopic, time-independent, active magnetic dipole can in principle exist without loss as a persistent DC current in a superconducting wire  loop or coil not requiring any extra energy or power input. For this situation all parts of Faraday's law in eqn. (\ref{Faraday}) are zero, as there is no voltage or emf required. The strength of the magnetic dipole depends on the enclosed electrical current in the loop. For a superconducting coil, a current may be trapped with the use of a persistent switch, and the strength of the magnetic field will depend on the applied $mmf$, $\mathcal{F}_T=NI$ before switching, as given by Ampere's law in eqn. (\ref{mmf2}). Thus, once trapped the $mmf$ exists as stored energy, $E_m=\frac{1}{2}LI^2$ ($L$ is the inductance of the loop or coil), and no work is required to keep the dipole energised.

The electromagnetic dual of the active macroscopic magnetic dipole (or permanent magnet) is the active macroscopic electric dipole. This type of dipole is a permanent dipole, such as a macroscopic electret, polar molecule, or atomic system with a dipole moment determined by the first-order linear Stark effect. This description does not include instantaneous or induced dipoles, which are not permanent. However, for the macroscopic electric dipole (or an electret) to exist an $emf$ must be generated to force separation of charges, unlike the magnetic dipole, this charge separation requires an impressed force per unit charge from an external energy source. For example, a solar cell contains a p-n junction, where an array of bound dipoles existed in the depletion region and when photons enter this region the dipoles are polarized to essentially form an electret and an electromotive force. Conversely, once an electret is polarized, the natural tendency is for the active electric dipole to discharge or decay and emit a photon \cite{Griffiths2011}, which means the active electric dipole is intrinsically metastable and are less common in nature. At the atomic scale a non-vanishing electric dipole moment is a much more rare occurrence than a magnetic dipole, which all particles with spin exhibit. By definition a non-vanishing electric dipole moment is proportional to a non-vanishing first-order Stark shift, which only occurs if some of the wavefunctions with degenerate energies have opposite parity; i.e., have different symmetry under inversion. This what happens for the excited H-atom, where 2s and 2p states are ``accidentally" degenerate and have opposite parity (2s is even and 2p is odd). In this paper we have presented a semiclassical emergent macroscopic description of EMF generation, where the voltage supplied by the active macroscopic electric dipole is determined by the enclosed effective magnetic current at the tangential boundary given by eqn. (\ref{Faraday}). In this dual system, the electric vector potential exists, and has a geometric phase. 

An interesting point in understanding the physics of an emf generator is to understand the microscopic description, which will be a different description for each type of generator, which inevitably involves quantum mechanics \cite{Jenkins2020,Alicki2020,Alicki2019,Hwang2012,LIU2020100066,Ilan:2020to} or a non-trivial microscopic material topology \cite{7820047}. Our work unifies this description with a simple emergent macroscopic description involving the modification of the Maxwell-Farady law. Another related question, is can we devise an experiment to measure the electric geometric phase in a similar way to the well-known AB experiment, which measures the magnetic geometric phase? Any experiment will need a full quantum mechanical description to understand if it would work, and act on the interference fringes of a passing particle such as an electron or a particle with an electric or magnetic dipole moment \cite{Chen2013,Dowling1999}. From fig. \ref{AR1}, we notice the vector potential is maximum just outside the rim of the dipole at the centre, at this same place the electric field is minimum. Passing particles around different directions would be the dual of the original AB experiment. Another way would be to configure an experiment which generates $emf$ in the regime dominated by the electric vector potential, and confirm the voltage output, this has already been undertaken with energy harvesters and Lorentz force generators, where electricity is generated by a bound or free charge polarisation in the absence of an applied electric field \cite{Tobar2021}.

\subsection*{Acknowledgements}

This work was funded by the Australian Research Council Centre of Excellence for Engineered Quantum Systems, CE170100009 and  Centre of Excellence for Dark Matter Particle Physics, CE200100008. We also thank Ian McArthur, Jay Sharping, Nabin Raul, Jeffery Miller, Walt Fitelson and David Mathes for participating in some discussions of this work.

\end{document}